\documentclass[journal]{IEEEtran}

\usepackage{amsmath}
\usepackage{graphicx}
\usepackage{afterpage}
\usepackage{ifthen}
\usepackage{amsmath}
\usepackage{theorem}
\usepackage{xspace}
\usepackage{calc}
\usepackage{amsfonts}
\usepackage{multirow}
\usepackage{subfigure}
\usepackage{amsfonts}
\usepackage{algorithm}
\usepackage{algorithmic}
\usepackage{setspace}

\usepackage[latin1]{inputenc}

\newcommand{\ie}{{\textit{i.e.},}\@\xspace}

\newcommand{\eg}{{\textit{e.g.},}\@\xspace}

\newcommand{\myvar}{{x}}
\newcommand{\myactvar}{{y}}
\newcommand{\mynstates}{{N_{s}}}
\newcommand{\ggdef}{\mathop {=} \limits^{\text{def}}}
\newcommand{\mygop}[1]{\myvar[#1]}
\newcommand{\myactgop}[1]{\myactvar[#1]}
\newcommand{\myframe}[2]{\myvar_{#1}[#2]}
\newcommand{\mynframe}{N_{f}}
\newcommand{\myngop}{N_{\text{GOP}}}
\newcommand{\mynview}{N_{\text{View}}}
\newcommand{\mypmf}{\textit{pmf}}

\newcommand{\myswitchpdf}[2]{d^{s}_{#2}[#1]}
\newcommand{\myavtime}[1]{\mu_{#1}}
\newcommand{\mystdtime}[1]{\sigma_{#1}}

\newcommand{\mypin}[1]{\pi_{#1}}
\newcommand{\myptr}[2]{\pi_{#1#2}}
\newcommand{\myPtr}{\Pi}
\newcommand{\mypmfs}[2]{b_{#2}[#1]}
\newcommand{\mypoiss}[2]{d_{#2}[#1]}
\newcommand{\mypoisscdf}[2]{D_{#2}[#1]}
\newcommand{\mydotpoiss}[2]{\dot{d}_{#2}[#1]}
\newcommand{\goplen}{N}
\newcommand{\myseq}{\myactvar_0^{N-1}}
\newcommand{\mypar}{\Theta}
\newcommand{\myparsp}{\mathbf{\Theta}}
\newcommand{\poisspar}{\lambda}
\newcommand{\Sseq}{S}
\newcommand{\Exp}{E}

\newcommand{\hlength}{8.2cm}
\newcommand{\myalfad}[3]{\alpha_{#3}(#1,#2)}
\newcommand{\myalfa}[2]{\alpha_{#2}(#1)}

\newcommand{\mygammad}[3]{\gamma_{#3}(#1,#2)}

\newcommand{\myxid}[4]{\xi_{#4}(#1,#2,#3)}
\newcommand{\mykron}[2]{\delta_{#1}^{#2}}

\newcommand{\mytab}[1]{\begin{table}[#1]}
\newcommand{\mytabend}{\end{table}}
\newcommand{\dummy}{0}
\newcommand{\quality}{0}
\newcommand{\mychr}{$c_r$}

\newcommand{\mysuballfig}[2]
{
\ifthenelse{#1 = 4}{\renewcommand{\dummy}{\ref{fig:gop4}}}{\renewcommand{\dummy}{\ref{fig:gop8}}}
\begin{figure*}[t!]
\centering
\subfigure[Autocorrelation function]{\includegraphics[width =\hlength, keepaspectratio]{./acf_G#1_Q#2_viewall.eps}}
\subfigure[ Q-Q Plot]{\includegraphics[width =\hlength, keepaspectratio]{./qq_G#1_Q#2_viewall.eps}}
\caption{Comparison of (a) the  autocorrelation function  and (b) the  Q-Q plot estimated on the real MVC encoded sequence  and on the synthetic P-HMM generated video sequence (GOP structure in Figure \dummy, \quality quality stream). \label{fig:sta#1-#2}}
\end{figure*}
}

\newcommand{\myallfig}[2]{
\ifthenelse{#1 = 4}{\renewcommand{\dummy}{\ref{fig:gop4}}}{\renewcommand{\dummy}{\ref{fig:gop8}}}
\ifthenelse{#2 = 10}{\renewcommand{\quality}{high~}}{\ifthenelse{#2 = 20}{\renewcommand{\quality}{medium~}}{\renewcommand{\quality}{low~}}}
\mysuballfig{#1}{#2}
}

\newcommand{\mybuft}{$B_T\ $}

\newcommand{\mybuftsize}{$b_T\ $}

\newcommand{\mycr}{\textit{r}}
\newcommand{\mydelay}{\textit{D}}
\newcommand{\myovdel}[1]{d[#1]}
\newcommand{\mycdel}[1]{d_C[#1]}
\newcommand{\mytdel}[1]{d_T[#1]}

\title{A Poisson Hidden Markov Model for Multiview Video Traffic}

\author{Lorenzo~Rossi, Jacob~Chakareski, Pascal~Frossard, and Stefania~Colonnese%
\thanks{L. Rossi and S. Colonnese are with
    DIET Dept.,
     Sapienza Universit\`a di  Roma, via Eudossiana 18, I-00184 Roma, Italy}%
\thanks{J. Chakareski and P. Frossard are with
 Signal Processing Laboratory (LTS4),
         Ecole Polytechnique F\'{e}d\'{e}rale de Lausanne (EPFL), CH-1015 Lausanne, Switzerland}
}

\begin{document}

\maketitle
\begin{abstract}
Multiview video has recently emerged as a means to improve user experience in novel multimedia services. We propose a new stochastic model to characterize the traffic generated by a Multiview Video Coding (MVC) variable bit rate source.  To this aim, we resort to  a Poisson Hidden Markov Model (P-HMM), in which the first (hidden) layer represents the evolution of the video activity and the second layer represents the frame sizes of the multiple encoded views. We propose a method for estimating the model parameters in long MVC sequences. We then present extensive numerical simulations assessing the model's ability to produce traffic with realistic characteristics for a general class of MVC sequences.
We then extend our framework to network applications where we show that our model is able to accurately describe the sender and receiver buffers behavior in MVC transmission. Finally, we derive a model of user behavior for interactive view selection, which, in conjunction with our traffic model, is able to accurately predict actual network load in interactive multiview services.
\end{abstract}
\section{Introduction}
\label{sec:intro}
The advent of novel video services with multiple views of the same video scene, e.g., 3-D TV or free-view point video, poses many novel challenges in terms of coding, processing and transmission of the multimedia content. As far as encoding techniques are concerned, the ISO/ITU-T Joint Video Team has recently finalized the H.264 Multiview Video Coding (MVC) standard, which is explicitly devoted to efficient compression of  a multiview source \cite{bib:mvc}. It is expected that multiview  video communication services will be traffic intensive, which raises important questions in network dimensioning. In addition, the  encoding dependencies between the different views renders resource allocation quite challenging in a MVC communication system.

Both problems of network dimensioning and resource allocation are usually addressed with the help of traffic models in classical video delivery services. Such tools have proved to be a valid support for efficient and accurate allocation of network resources by characterizing the compressed video content through statistical models.
Video traffic models have been derived for different applications in teleconferencing \cite{bib:tlc}, video broadcasting \cite{bib:liu1,bib:len}, or streaming \cite{bib:isivc}. Different stochastic models based on autoregressive processes \cite{bib:tlc}, Transform Expanded Sample (TES) processes \cite{bib:tes}, and Hidden Markov Models (HMMs) \cite{bib:eus} have been considered for network design, resource allocation, buffer dimensioning, and performance evaluation \cite{bib:cac}.

There is however a lack of traffic models for multiview video communication services. We propose here a new traffic model for MVC content in order to characterize the frame size sequence observed at the output of an MVC  variable bit rate (VBR) source.  Specifically, building on our preliminary work  \cite{bib:icip10}, we design a doubly stochastic source model, namely a Poisson Hidden Markov Model (P-HMM) \cite{bib:pois}, in which the first (hidden) layer consists of a non-stationary chain modeling the video activity level and the second layer represents the frame sizes of the different MVC encoded views. Besides, we extend the P-HMM parameter   estimation algorithm for short observation sequences  presented in \cite{bib:pois} and adapt it to long sequences such as those encountered in video communication services. We assess the model performances by extensive numerical simulations on classes of MVC sequences sharing common properties.
We apply our model to predict the traffic load generated by two different network services based on a client-server video communication paradigm.  In the first one, that we name Multiview TV, the server simultaneously streams all the MVC encoded views to the client. Our model is shown to be able to predict the state of sender and receiver buffers in Multiview TV. In the second one, named interactive TV, the client dynamically  selects the views  during the streaming session by means of a feedback channel. Due to the MVC encoding dependencies, the server transmits a composite stream encompassing all encoded data required to correctly decode the selected view.
Finally, we introduce an Interactive TV  user service request model, in order to mimic  the sequence of requested views selected by the user. In fact, the traffic generated during the interactive TV session depends both on the MVC encoded video traces and on the user's view selection. We show that the combination of our two models is able to accurately characterize the traffic in interactive multiview applications. 

The main contribution of this paper can be summarized as follows:
\begin{itemize}
\item a non-stationary traffic model for VBR MVC sequences is introduced, with the ability to characterize different classes of MVC streams at different encoding settings. The model can predict actual network load in network applications;
\item a Maximum Likelihood (ML) estimation procedure suitable to derive the traffic model parameters in long sequences is derived;
\item a user behavior model for interactive view selection is combined with our traffic model to characterize interactive multiview traffic.
\end{itemize}

The rest of the paper is organized as follows. In Section \ref{sec:phmm}, we introduce the Poisson Hidden Markov Model (P-HMM) and discuss its feasibility; we also describe the P-HMM parameter set estimation procedure. In Section \ref{sec:mas}, we validate the model in different stream settings. Network applications of our model are studied in Section \ref{sec:buf}, along with the view switching model. Section \ref{sec:conc} concludes the paper.

\section{MVC Source Modeling}
\label{sec:phmm}
\subsection{MVC coding format}
\label{sec:code}
\begin{figure}[h]
\centering
\renewcommand{\hlength}{8cm}
\includegraphics[width=\hlength, keepaspectratio]{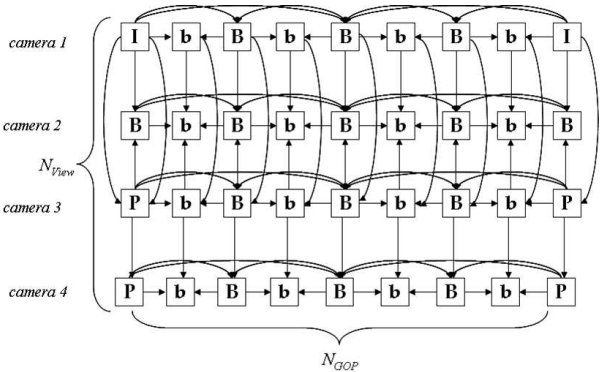}
\caption{\label{fig:gop8}GOP structure and encoding hierarchy  (\protect$\myngop=8$).}
\end{figure}

\begin{figure}[h]
\centering
\renewcommand{\hlength}{5.5cm}
\includegraphics[width=\hlength, keepaspectratio]{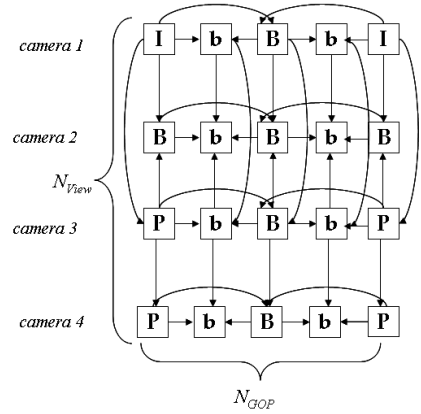}
\caption{\label{fig:gop4}GOP structure and encoding hierarchy   (\protect$\myngop=4$).}
\end{figure}

A MVC stream jointly encodes different video sequences captured  by multiple cameras with overlapping fields of view. Let us denote by $\mynview$ the number of such sequences. One view, denoted as reference view, is independently encoded using temporal motion compensation and transform coding techniques, similarly to a classical video sequence encoded with the H.264 encoder \cite{bib:mvc}. The other $\mynview-1$ views are encoded using inter-view prediction in addiction to temporal prediction, in order to further improve the compression performance. In H.264 MVC \cite{bib:mvc}, inter-view prediction is allowed between frames referring to the same time instant, whereas intra-view encoding  dependencies are usually set to permit temporal scalability \cite{bib:svc}.  The encoding dependencies  give rise to a generalized Groups Of Pictures (GOP) structure of duration $\myngop$, constituted by $\mynframe = \mynview \times \myngop$ frames. Figures \ref{fig:gop8} and \ref{fig:gop4} show two examples of such MVC GOPs. Given the complete  MVC encoded bitstream, up to  $\mynview$ flows are transmitted and decoded by the client. In most applications, all the views are transmitted together in a simulcast mode.

In order to control the size of the bitstreams, different rate control algorithms \cite{bib:mvcrc} can be implemented in MVC encoders. We however focus in this paper on Variable Bit Rate (VBR) streams, where the quantization step sizes are fixed. Their value only depends on the frame type, as commonly employed in most MVC applications \cite{bib:mvc}.

\subsection{Traffic model}
\label{sec:tm}
We propose now a new model that is able to characterize the frame sizes  in GOPs of an MVC compressed stream with a given GOP structure. In VBR operating mode,  the bit-rate of the MVC encoded views  varies according to the video activity level, and the traffic model should match this non stationary stochastic process. We propose a two-layer stochastic process model inspired from \cite{bib:Doulamis}. We build a new Poisson Hidden Markov Model (P-HMM) \cite{bib:pois}, in which the first (hidden) layer is a discrete time Markov chain whose states represent different video activity levels; the second layer represents the frame size sequence corresponding to a given activity level. We make the following assumptions about the non-stationarity of the sources:
\begin{enumerate}
\item the  activity level of the video content varies in time according to a Poisson distribution;
\item the hidden layer state transitions, (\ie the change of the activity level) occur at the beginning of a GOP;
\item the activity level is the same in all views since views are correlated.
\end{enumerate}

With these simplifying assumptions, we approximate the duration of scenes with a given activity level, by a simple one parameter distribution. We discard the possible changes in the average frame size due to activity level changes observed in the middle of a GOP. We will show that, in spite of this approximation, the model closely matches the MVC source characteristics.

Let us denote the number of states (\ie different video activity levels) in our model as $\mynstates$.
The state duration obeys a  Poisson distribution  denoted by
\[\mypoiss{k}{i} = \frac{e^{-\poisspar_i}\poisspar_i^k}{k!}.\]
When a state transition occurs, the model is described by a state transition matrix $\myPtr$, whose element $\myptr{i}{j}$  denotes the probability of transition from state $i$ to state $j$. Because of the explicit modeling of the state duration time distribution, $\myptr{i}{i} = 0$ for all $i$. Finally, $\pi_i$ denotes the initial probability of the model being in state $i$.

The second layer in our model describes the frame sizes in an entire GOP. Formally, let us consider the random vector \[\mygop{n} = [\myframe{0}{n}, \dots, \myframe{\mynframe-1}{n}]\] representing the set of frames sizes in the $n$-th GOP of the compressed multiview content.
The vector   $\mygop{n}$ is emitted in accordance to a multivariate probability mass function (\mypmf) depending on the actual hidden layer state.
Given the current state in the first layer of the model, say $i$, a random vector $\mygop{n}$ is generated according to the \mypmf ~$b_i[\mygop{n}]$. For the sake of compactness, each \mypmf ~$b_i[\cdot]$, $i=1,\dots,\mynstates$, has a different number of bins depending on the coding mode (namely I, P, or B-frames) of the compressed picture to be generated. This choice is motivated by not imposing that the frame size \mypmf~ follows a fixed probability distribution (\eg gaussian, gamma, \dots), but instead by adapting the distribution shape to the actual MVC sequences. Moreover, by allowing a different number of bins for different frame types, we can adaptively control the model's complexity (\ie the parameter set) according to the desired accuracy performance.

\subsection{Parameter estimation}
\label{sec:EM}
The estimation of the model parameters is a crucial step in traffic characterization. Since the model belongs to the wide HMM family \cite{bib:rab}, we can resort to one of the estimation algorithms employed for such models. In particular, an estimation procedure called Expectation-Maximization (EM) algorithm \cite{bib:ema} is widely used  for HMMs. A version of the EM algorithm has been proposed for P-HMMs in \cite{bib:pois}. However, it exhibits numerical instability when used for long data sequences \cite{bib:eph}. We derive here a new EM algorithm for stable parameter estimation in long sequences. Our algorithm extends the method of \cite{bib:eph} to the case of non-stationary hidden state durations. We present the parameter estimation in detail below.

Suppose that we observe a video sequence composed of $\goplen$ GOPs. Let $\myseq \ggdef \{ \myactgop{n} \}_{n=0}^{N-1} $ denote the observed video traffic and $\mypar  \in  \myparsp $ the parameter set of our model, where $\myparsp$ is the parameter space and $\mypar \ggdef \{\myPtr, \poisspar_1, \mypmfs{\cdot}{1},\mypin{1}, \dots, \poisspar_\mynstates, \mypmfs{\cdot}{\mynstates},\mypin{\mynstates} \}$. The  EM  algorithm  comprises two iterative computational steps. The first one is an expectation step that computes the auxiliary  likelihood function $Q(\mypar|\mypar^{(m)}) = \Exp\{\log(\text{Prob}\{\Sseq,\myvar,\mypar\})|\myactvar,\mypar^{(m)}\}$, in which $\Sseq \in \mathbf{\Sseq}$ represents a plausible state sequence and $\mypar^{(m)}$ is the $m$-th estimate of the parameter set. Then, a maximization step maximizes the likelihood function, \ie
\begin{equation}
\mypar^{(m+1)} = \underset{\mypar}{\operatorname{arg\,max}} \, Q(\mypar|\mypar^{(m)}).
\end{equation}
The algorithm iterates between the two steps until convergence of the parameter set.

Before getting into more details, let us define a function $\mydotpoiss{k}{i}$ representing the state duration distribution and taking into account the finite length of the observed sequence, as
\begin{equation}
\mydotpoiss{k}{i} =
\begin{cases}
1-\mypoisscdf{k-1}{i} &\text{if} \;\;k = \goplen - n - 1\\[10pt]
\mypoiss{k}{i} &\text{otherwise}
\end{cases}.
\end{equation}
The parameter $\mypoisscdf{k}{i}$ is the cumulative distribution of the state duration times. Then, the computations in our EM algorithm, as applied to P-HMMs, comprise i) the computation of forward probabilities, ii) the computation of backward probabilities, iii) the estimation of the parameter set.
The first two steps calculate three auxiliary variables, namely the conditioned probabilities of the state sequence to the observed sequence, representing the expectation steps of our EM algorithm (see \cite{bib:rab} for further details). Then, the parameter set is expressed as function of these auxilary variables. We first define the following forward probabilities \footnote{Our definitions differ from \cite{bib:pois} in order to avoid numerical instability.} for $n = 0, \dots, \goplen-1$
\begin{align}
&\begin{aligned}
\myalfad{i}{k}{n} &\ggdef P(s_n = i,\dots,s_{n+k} = i,\\
& s_{n+k+1} \ne i | \myactvar_0^n,\mypar^{(m)}),&k <\goplen-n-1\end{aligned}\label{eq:alfad}\\[4pt]
&\myalfa{i}{n} \ggdef P(s_n = i| \myactvar_0^n,\mypar^{(m)})\label{eq:alfa}.
\end{align}
For $k = \goplen-n-1$ the definitions above are slightly different in order to  take into account the finite length of the actual sequence
\begin{equation*}
\myalfad{i}{\goplen-n-1}{n} \ggdef P(s_n = i,\dots,s_{\goplen-1} = i | \myactvar_0^n,\mypar^{(m)}).
\end{equation*}
Those quantities are calculated by the recursive algorithm illustrated in Algorithm \ref{tab:alfa}\footnote{The normalization coefficient for $\myalfad{i}{k}{n}$ is calculated by summation over $i$ and $k$.}. They represent the probability for the system to be in state $i$ at time $n$ and to stay in the same state for the next $k$ instants, given the sequence observed till time $n$.

\begin{algorithm}
\begin{spacing}{1.5}
\begin{algorithmic}[1]
\FOR{$n = 0 \to \goplen-1$}
\FOR{$k = 0 \to \goplen-n-1$}
\IF{$n = 0$}
\STATE
\(
\myalfad{i}{k}{0} \propto \mypin{i}\mydotpoiss{k}{i}\mypmfs{\myactgop{0}}{i}
\)
\ELSE
\STATE \(
\myalfad{i}{k}{n} \propto\  \mypmfs{\myactgop{n}}{i}(\sum_{\substack{j = 1\\j \ne i}}^\mynstates \myalfad{j}{0}{n-1}\myptr{j}{i}\mydotpoiss{k}{i}
+\myalfad{i}{k+1}{n-1}) \)
\ENDIF
\ENDFOR
\STATE \(\myalfa{i}{n} = \sum_k \myalfad{i}{k}{n}\)
\ENDFOR

\end{algorithmic}
\end{spacing}
\caption{\label{tab:alfa}Computing forward probabilities $\myalfad{i}{k}{n}$ and $\myalfa{i}{n}$.}

\end{algorithm}

Then, the following backward probabilities are defined in a similar way 

\begin{align}
&\begin{aligned}
\mygammad{i}{k}{n} &\ggdef P(s_n = i,\\
&\dots,s_{n+k} = i, s_{n+k+1} \ne i | \myseq,\mypar^{(m)}),\\
& n = 0, \dots, \goplen - 1;\ k < \goplen - n - 1\end{aligned}\label{eq:gammad}\\[10pt]
&\begin{aligned}
\myxid{i}{j}{k}{n} &\ggdef P(s_{n-1} =i,\ s_{n} = j,\\
&\dots,s_{n+k} = j, s_{n+k+1} \ne j  | \myseq,\mypar^{(m)}),\\
& n = 1, \dots, \goplen - 1;\ k < \goplen - n - 1.\label{eq:xid}
\end{aligned}
\end{align}
Note that we resort to a different definition for the backward probabilities with respect to the usual $\beta$~notation \cite{bib:rab}. In \cite{bib:rab},  a $\beta$~backward probability is defined representing the likelihood of the observed sequence, and $\gamma$~and $\xi$~are calculated by means of $\alpha$~and $\beta$. Here, we calculate directly $\gamma$~and $\xi$~ in a backward iteration in order to avoid numerical issues arising from the sequence length.
The Algorithm \ref{tab:xi} illustrates the backward probabilities computation, where $\mykron{i}{j}$ denotes the Kronecker function.
\begin{algorithm}
\begin{spacing}{1.5}
\begin{algorithmic}[1]
\FOR{$n = \goplen-1 \to 1$}
\FOR{$k = \goplen-n-1 \to 0$}
\IF{$n = \goplen-1$}
\STATE
\(
\mygammad{i}{k}{\goplen-1} = \myalfa{i}{\goplen-1}
\)
\ELSE
\IF{$k \ne 0$}
\STATE $\mygammad{i}{k}{n} =\myxid{i}{i}{k-1}{n+1}$
\ELSE
\STATE $\mygammad{i}{0}{n} =\sum_{\substack{j = 1\\j \ne i}}^{\mynstates}\sum_{k=0}^{\goplen-n-2}\myxid{i}{j}{k}{n+1}$
\ENDIF
\ENDIF
\STATE \(\myxid{i}{j}{k}{n} = \frac{\myalfad{i}{0}{n-1}\myptr{i}{j}\mydotpoiss{k}{j}+ \mykron{i}{j}\myalfad{i}{k+1}{n-1}}{\sum_{l = 1}^{\mynstates} \myalfad{l}{0}{n-1}\myptr{l}{j}\mydotpoiss{k}{j}+ \mykron{l}{j}\myalfad{l}{k+1}{n-1}}\cdot\mygammad{j}{k}{n}\)
\ENDFOR
\ENDFOR
\end{algorithmic}
\end{spacing}
\caption{\label{tab:xi}Stable estimation of backward probabilities $\mygammad{i}{k}{n}$ and $\myxid{i}{j}{k}{n}$, from \eqref{eq:gammad}-\eqref{eq:xid}}
\end{algorithm}

Finally, the parameter set $\mypar^{(m+1)}$ in the maximization step can be calculated with the help of the forward and the backward probabilities above. We can write the initial probability of being in state $i$ as:
\begin{equation}
\mypin{i} = \sum_{k=0}^{\goplen-1}\mygammad{i}{k}{0}.
\label{eq:startp}
\end{equation}
Then we can express the transition probabilities as:
\begin{equation}
\myptr{i}{j}
= \frac{\sum_{n=1}^{\goplen-1}\sum_{k=0}^{\goplen-n-1}\myxid{i}{j}{k}{n}}{\sum_{\substack{j = 1\\j \ne i}}^{\mynstates}\sum_{n=1}^{\goplen-1}\sum_{k=0}^{\goplen-n-1}\myxid{i}{j}{k}{n}}.
\end{equation}
The frame size distribution in each state is given by:
\begin{equation}
\mypmfs{\myvar}{i} = \frac{\sum_{n=0}^{\goplen-1}\sum_{k=0}^{\goplen-n-1}\mygammad{i}{k}{n}\mykron{\myvar}{\myactgop{n}}}{\sum_{n=0}^{\goplen-1}\sum_{k=0}^{\goplen-n-1}\mygammad{i}{k}{n}}.
\end{equation}
Finally, the state duration is expressed as:
\begin{multline}
\hspace{-0em} \lambda_{i} = \left(\sum_{n=1}^{\goplen-1}\sum_{k=0}^{\goplen-n-2}\sum_{\substack{j = 1\\j \ne i}}^{\mynstates}k\ \myxid{j}{i}{k}{n}+\sum_{k=0}^{\goplen-1}k\ \mygammad{i}{k}{0}\right)\\
\cdot\left(\sum_{n=1}^{\goplen-1}\sum_{k=0}^{\goplen-n-2}\sum_{\substack{j = 1\\j \ne i}}^{\mynstates}\myxid{j}{i}{k}{n}+\sum_{k=0}^{\goplen-1}\mygammad{i}{k}{0}\right)^{-1}.
\label{eq:lambda}
\end{multline}
Note that a single iteration of the estimation algorithm consists of calculating first the state sequence probabilities \eqref{eq:alfad}-\eqref{eq:xid}, which is the expectation step. Subsequently, we compute the new parameter estimates by averaging the observations weighted with the state probabilities \eqref{eq:startp}-\eqref{eq:lambda}, which corresponds to the maximization step of an iteration of the algorithm. Convergence is assured by Jensen's inequality \cite{bib:eph}.

\section{Traffic model validation}
\label{sec:mas}
In this section, we assess our model by comparing statistics evaluated on a pseudo-random synthetic traffic generated according to our P-HMM, with the statistics evaluated on a composite MVC test sequence. The P-HMM parameters are estimated by applying the EM algorithm of Section \ref{sec:EM} on the observed composite sequence. We also test the case in which the test sequence is different from the sequence used to train the model, in order to show the model ability to represent not only the training sequence, but also other sequences with similar content. A comparison with a well-known single view VBR model is also carried out.

\subsection{MVC encoder settings}
\label{sec:simset}
The composite MVC sequences considered in the model assessment are generated by concatenating several  tests sequences with very different activity levels, as reported in Table \ref{tab:seq}. All the sequences are in CIF format, with a frame rate of 25 fps,  and $\mynview = 4$ views. The resulting MVC composite sequence  is approximately 6 minutes long.  We encode the  sequence using the two different GOP structures reported in Figures \ref{fig:gop8} and \ref{fig:gop4}. Both structures exhibit motion compensation dependencies among the views for anchor and non-anchor frames \cite{bib:gop}. The large number of dependencies accentuates the difference between MVC traffic and simple aggregations of single view traffic. The two GOP structures differ in the number of I and P frames. 
The bit-rate variability of sequences encoded using the GOP in Figure \ref{fig:gop8} is mainly due to the residuals of the motion compensation whereas for sequences encoded using the GOP in Figure \ref{fig:gop4}; the bit-rate variability depends on the large number of intra or P frames.
For each GOP structure, different MVC bitstreams have been generated by setting the quantization parameter of the reference view to 10 (high quality), 20 (medium quality), or 40 (low quality), and by adjusting the temporal layers quantization parameter accordingly%
\footnote{
Specifically, we have set 
the quantization parameters according to the default settings of the JMVC \cite{bib:jmvc}.
}. %
We use  JMVC v7.0 to encode the sequences \cite{bib:jmvc}, then we use the different MVC bitstreams to build traffic models.
\begin{table}[!htb]
\center
\begin{tabular}{|c|c|}
\hline
\textbf{sequence name} & \textbf{\# frames} \\
\hline
Akko \& Kayo & 290 \\
\hline
Champagne Tower & 500\\
\hline
Uli & 250\\
\hline
Jungle & 250\\
\hline
Balloons & 500\\
\hline
Kendo & 400\\
\hline
Dog & 300\\
\hline
Pantomime & 500\\
\hline
\end{tabular}
\caption{ \label{tab:seq}Test sequences used to generate the compound sequence.}
\end{table}

As we discussed in the previous section, the number of bins in the \mypmf s of the model may be different for different frame types in order to have a trade-off between performance and model complexity. In our tests we use the number of bins shown in Table \ref{tab:binsss} \subref{tab:bins1} and \subref{tab:bins2} for  the GOP structures in Figures \ref{fig:gop8} and \ref{fig:gop4}, respectively. The bins are placed in the interval between the minimum and the maximum observed frame size for each frame type.

\begin{table}[h]
\centering
\subtable[\label{tab:bins1}GOP structure in Figure \ref{fig:gop8}.]{
\begin{tabular}{|l|c|c|c|c|c|c|c|c|}
\hline
View \#0 &50 &10 &10 &10 &20 &10 &10 &10\\
\hline
View \#1 to \#3 &30 &10 &10 &10 &20 &10 &10 &10\\
\hline
\end{tabular}}
\subtable[\label{tab:bins2}GOP structure in Figure \ref{fig:gop4}.]{
\begin{tabular}{|l|c|c|c|c|c c c c|}
\hline
View \#0 &50 &10 &20 &10\\
\hline
View \#1 to \#3 &30 &10 &20 &10\\
\hline
\end{tabular}}
\caption{\label{tab:binsss}Number of bins in the pmf for each frame of the GOP given in display order.}
\end{table}

In the simulations, we employ a 3-state model in order to represent low, medium and high level activity. A first coarse estimation of the model parameter is performed by labeling each GOP of the actual sequence as low, medium or high according to its average frame size. The thresholds are set in order to have the same number of GOPs in the three states. Then, a coarse estimation is obtained by evaluating frame size histograms related to each state. Zero-valued bins are set to a low fixed value and the histograms are normalized accordingly. The transition probability matrix is initializated with positive random values, and the Poisson mean values are set to 1 for each state. After that, the EM algorithm is performed as described in Section \ref{sec:EM} using the coarse estimation as starting point. Estimation ends when the difference between the log likelihood  of the two most recent iterates is smaller than 0.01. 

Finally, synthetic traffic is produced by first generating a state sequence according to the model and then producing synthetic traffic for a GOP for each state of the sequence by means of the frame size \mypmf s. The bins are converted to the mean frame size of the interval they represent. The synthetic traffic generation is described in Algorithm~\ref{alg:syn}.
\begin{algorithm}[h!]
\begin{spacing}{1.5}
\begin{algorithmic}[1]
\STATE Initial state $i$ extracted according to the distribution $\pi_1, \pi_2,\dots,\pi_{N_s}$
\STATE $n \gets 0$
\WHILE{$n < \goplen$}
\STATE $k$ extracted from the Poisson distribution belonging to the current state, i.e., $\mypoiss{k}{i}$
\IF {$n \ge \goplen - k$}
\STATE $k \gets \goplen - n - 1$
\ENDIF
\FOR{$j =  0\ \text{to}\ k$}
\STATE generate a synthetic GOP $\mygop{n + j}$ according to $b_i[\cdot]$;
\ENDFOR
\STATE {$n \gets n + k$}
\STATE perform state transition according to transition matrix $\myPtr$.
\ENDWHILE
\end{algorithmic}
\end{spacing}
\caption{\label{alg:syn}Synthetic traffic generation.}
\end{algorithm}
\subsection{Performance evaluation}
\myallfig{4}{10}
\myallfig{4}{20}
\renewcommand{\hlength}{8cm}
\renewcommand{\hlength}{8cm}
\begin{figure*}[thp!]
\subfigure[autocorrelation function]{\includegraphics[width =\hlength, keepaspectratio]{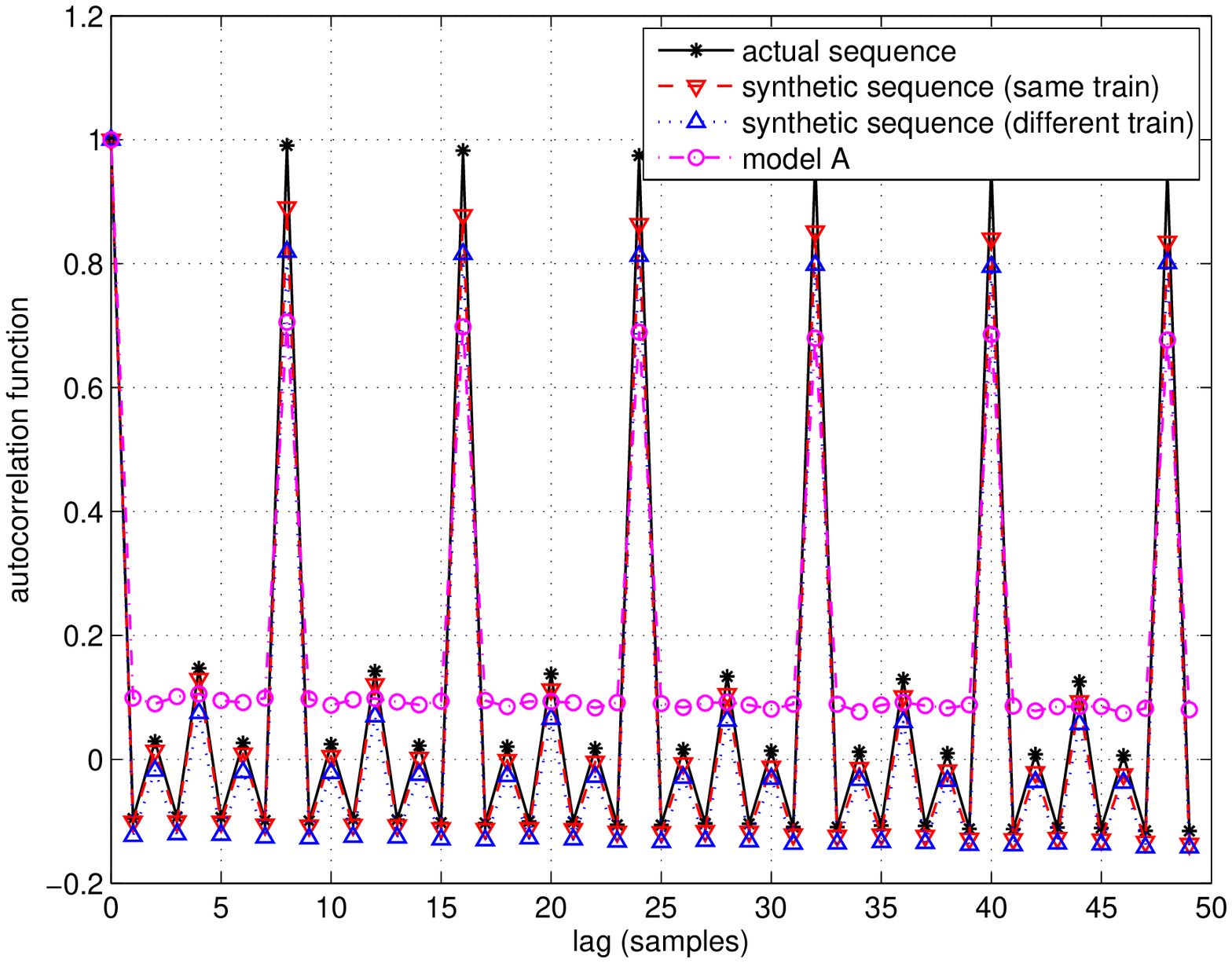}}
\subfigure[Q-Q plot]{\includegraphics[width =\hlength, keepaspectratio]{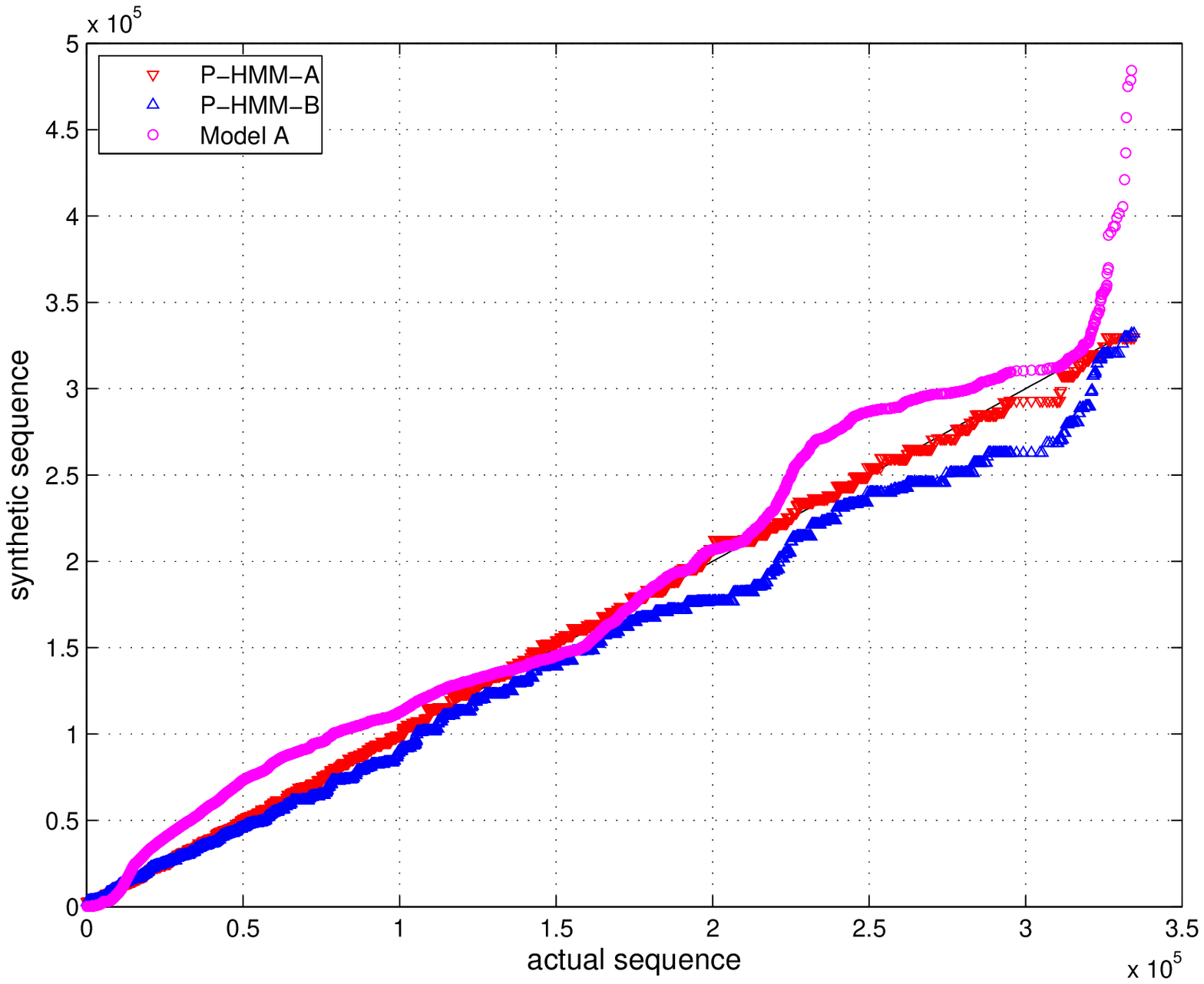}}
\caption{\label{fig:sta8-20} Q-Q plot and autocorrelation estimated on the actual MVC sequence and the synthetic sequences (GOP structure in Figure \ref{fig:gop8}, medium quality stream).}
\end{figure*}
\renewcommand{\hlength}{8cm}
\begin{figure*}[thp!]
\subfigure[autocorrelation function]{\includegraphics[width =\hlength, keepaspectratio]{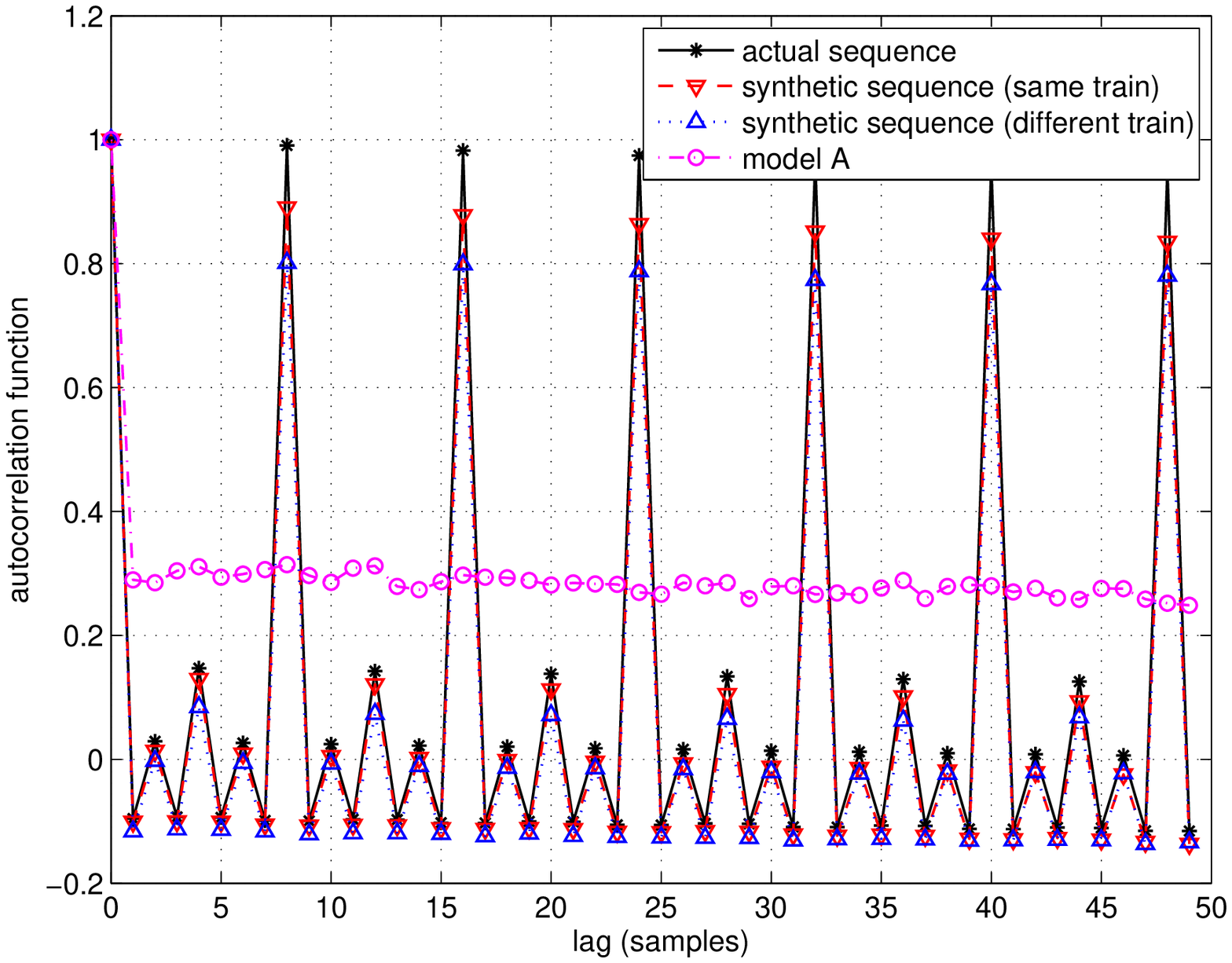}}
\subfigure[Q-Q plot]{\includegraphics[width =\hlength, keepaspectratio]{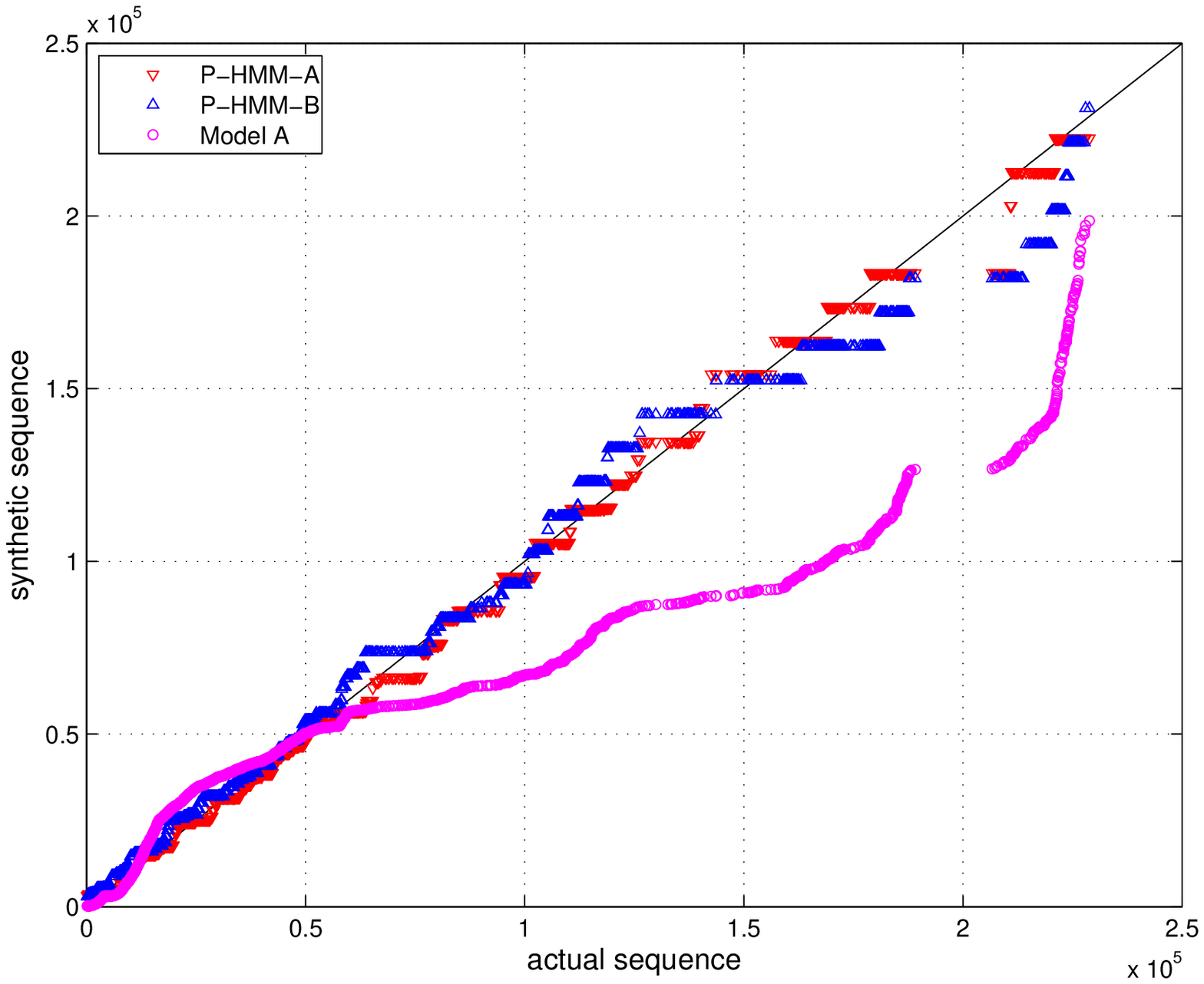}}
\caption{\label{fig:sta8-20-1} Q-Q plot and autocorrelation estimated on the actual MVC sequence (View \#1) and the synthetic sequences (GOP structure in Figure \ref{fig:gop8}, medium quality stream).}
\end{figure*}
\renewcommand{\hlength}{8cm}
\begin{figure*}[thp!]
\subfigure[autocorrelation function]{\includegraphics[width =\hlength, keepaspectratio]{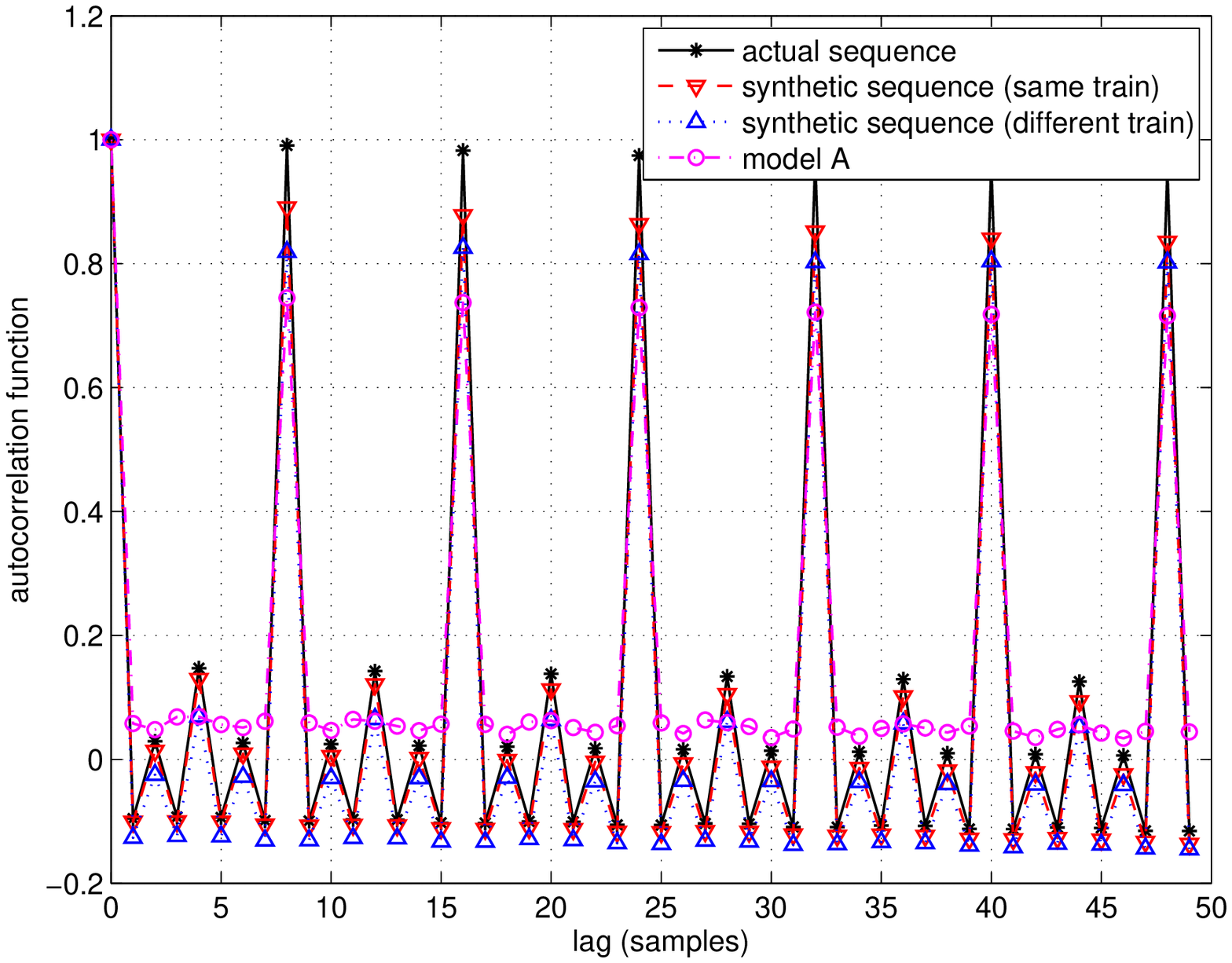}}
\subfigure[Q-Q plot]{\includegraphics[width =\hlength, keepaspectratio]{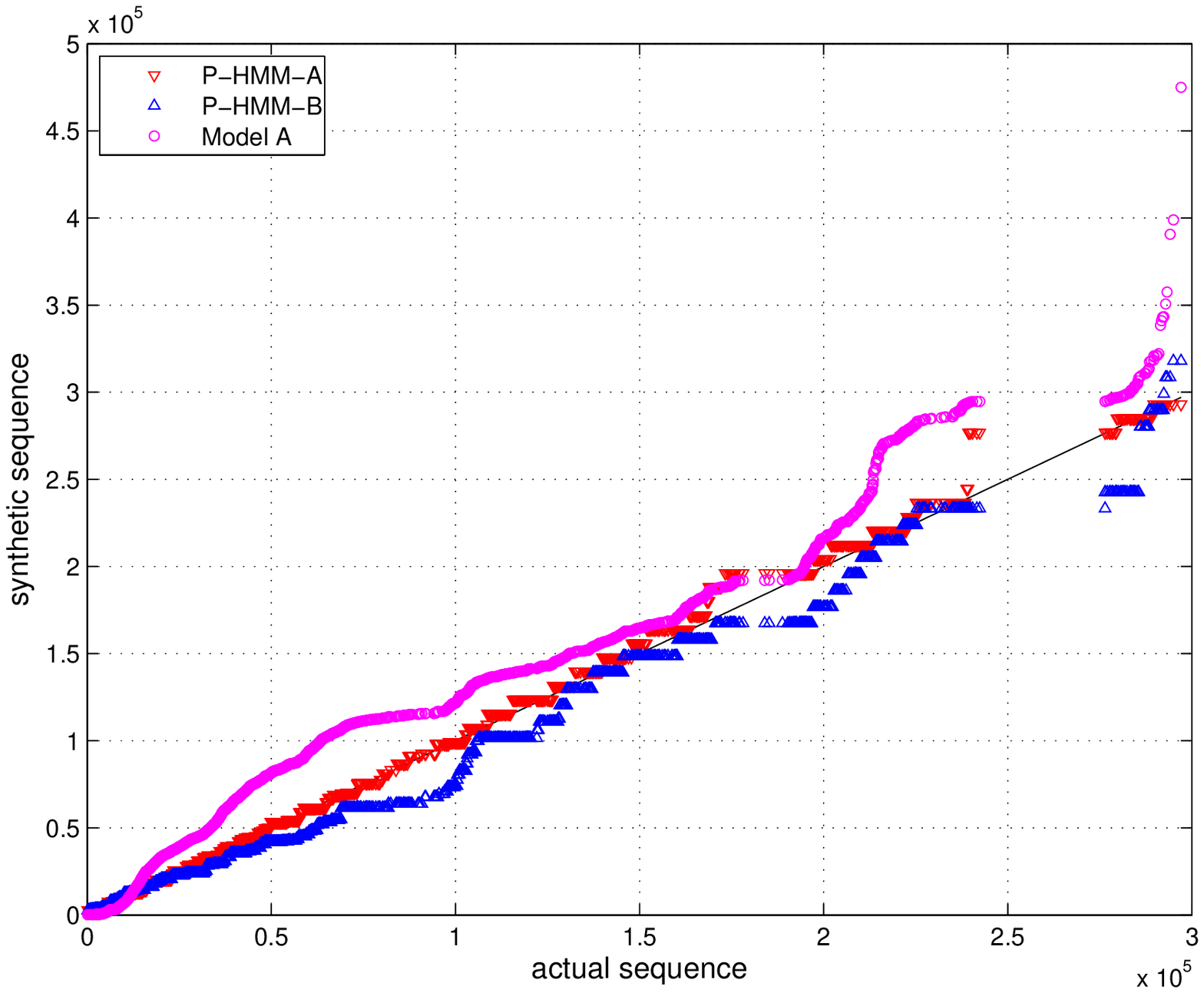}}
\caption{\label{fig:sta8-20-3} Q-Q plot and autocorrelation estimated on the actual MVC sequence (View \#3) and the synthetic sequences (GOP structure in Figure \ref{fig:gop8}, medium quality stream).}
\end{figure*}
We now assess the model's accuracy by first comparing  the autocorrelation function (acf) and the Q-Q 
plot computed on a sequence of frame sizes of the actual MVC encoded sequence (comprising all the views) with the respective statistics evaluated on a pseudo-random traffic sequence generated by the P-HMM with parameters  estimated from the corresponding composite sequence. 
Figures \ref{fig:sta4-10}, \ref{fig:sta4-20} and \ref{fig:sta8-20} show the  acf and the Q-Q plot for different compressed MVC sequences. Q-Q plot measures how two distributions are similar by comparing their quantiles. The closer the Q-Q plot to the bisector of the first and third quadrant, the more similar the two distributions are.It is clear that the statistics evaluated on the synthetic traffic (P-HMM-A) closely follow the statistics of the actual MVC sequences, for every GOP structure  and quantization parameter. The close match with the actual data is due to the fact that we do not constrain the frame size \mypmf s to have an explicit probability distribution and that we employ the Poisson distribution for the state durations, thus forcing a non-stationary recurrence of the activity levels. In order to validate these conclusions, we compare our model to the well-known single view VBR model described in \cite{bib:gamma}. Specifically, we focus on the "model A" in \cite{bib:gamma}, whose main differences with our P-HMM model relate to the frame size distribution and the hidden chain describing the activity level. The model A employs three shifted gamma distributions for the sizes of frames (I, P, B) and a stationary 7-state Markov chain for the activity level. Moreover, an ad-hoc parameter estimation procedure is used in \cite{bib:gamma}. We see in the Figures \ref{fig:sta4-10}, \ref{fig:sta4-20} and \ref{fig:sta8-20} that model A is not able to describe it correctly. In particular, we can see that model A does not depict accurately the shape of the actual sequence, both for the Q-Q plots and the autocorrelation function. We finally consider the case in which our model is trained with a different sequence with respect to the test sequence, we denote the synthetic traffic generated by this model as P-HMM-B. P-HMM-A outperforms the other models in mimicking the overall frame size distribution while P-HMM-B also achieves good adherence performance. 

The same statistics have been calculated separately on the traffic related to each view in MVC streams.  Figures \ref{fig:sta8-20-1}  and \ref{fig:sta8-20-3} show these statistics for the view \#1 and view \#3, respectively.  Similar results have been obtained for different encoder settings. Again, it is clear that model A \cite{bib:gamma} is not able to capture the statistics for each single view. Conversely,  our models
can characterize efficiently the traffic for each view. The characterization of the first and second order statistics for every view makes the model attractive to describe real MVC traffic in network applications, where a subset of the views are transmitted to the receivers. The good results obtained with PHMM-B show that the model is able to capture features that are not only related to a single MVC stream, but also to a class of sequences sharing similar content characteristics. Finally, we remark that the slight divergence in terms of autocorrelation function is caused by neglecting the intra-GOP correlation in the model design.

\renewcommand{\hlength}{4cm}

\section{Traffic model in multiview services}
\label{sec:buf}
\subsection{Applications scenarios}
We examine the accuracy of our model in the context of multiview services. We consider two case studies illustrated in Figure \ref{fig:sys}. In the first one, called "Multiview TV", the server sends all the MVC content to the user. In the second one, denoted as "Interactive TV",  the user requests one view at the time and can switch dynamically among the available views during the playout, exploiting an out-of-band feedback control channel. Due to the coding dependencies, the reference views still have to be transmitted along with the target view in the Interactive TV service. 

\renewcommand{\hlength}{9cm}
\begin{figure}[!htb]
\centering
\includegraphics[width=\hlength, keepaspectratio]{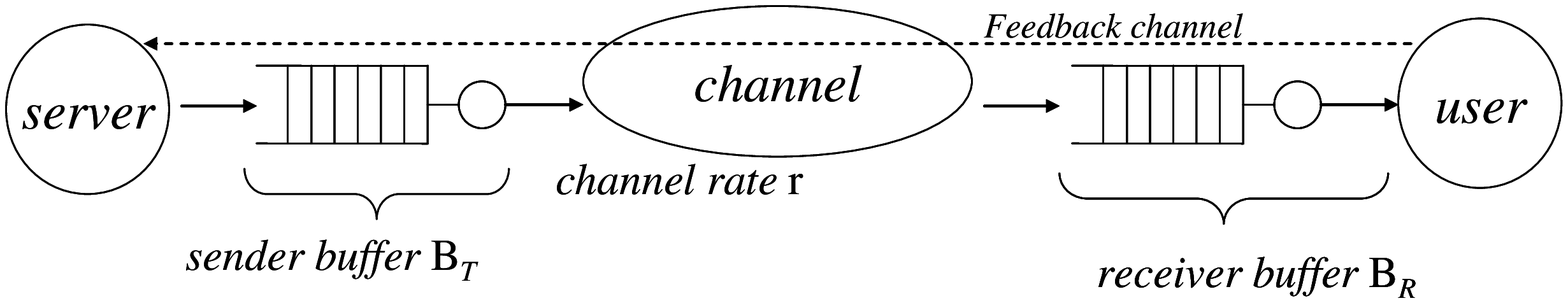}
\caption{System description for multiview services. The feedback channel is present only in the interactive TV case.}
\label{fig:sys}
\end{figure}
From the point of view of the MVC source model, the above services differ in that the traffic generated during the Multiview TV session depends only on the encoded video, whereas the traffic generated during the Interactive TV session depends also on the view selection process. Thereby, in order to fully characterize the latter scenario, we develop a new model of the view switching sequence, inspired by related models in the context of channel switching in IPTV systems \cite{bib:lee}-\cite{bib:cha}. 
In the design of our view switching model, we employ the following assumptions:
\begin{enumerate}
\item A user watches the reference view most of the time;
\item Other views are occasionally selected by the user because they represent added content with respect to the main view (\eg in a football game, these views may describe the foreground of the players);
\item A user selects views with preferences that depend on the present view the user is currently watching (first order dependency).
\end{enumerate}
We model the view switching sequence as a chain in which the states represent different views. The duration of stay in each view is explicitly modeled with a probability distribution. According to the above assumptions, we select sample values for the view transition probabilities, which are reported in Table \ref{tab:swmat} \subref{tab:tm}. The average and standard deviation of state durations are also set to sample values corresponding to the above assumptions (see Table \ref{tab:swmat} \subref{tab:sd}). We have found that the Gamma distribution is suitable and flexible for modeling the duration time, since it takes only non-negative values with mean and standard deviation that are features independent.  Formally, let $\myswitchpdf{t}{i}$ be the density function for the duration time in state $i$:
\begin{equation}
\myswitchpdf{k}{i} \ggdef \frac{\beta_{i}^{\alpha_{i}}}{\Gamma(\alpha_{i})} k^{\alpha_{i}-1} e^{-\beta_{i} {k}} \text{ for }k \geq 0
\end{equation}
The parameters $\alpha_{i}$ and $\beta_{i}$ are derived from the mean and the standard deviation of duration time for the corresponding state, respectively $\myavtime{i}$ and $\mystdtime{i}$, according to the following expressions:
\begin{equation}
\begin{cases}
&\alpha_{i} = \frac{\myavtime{i}^2}{\mystdtime{i}^2}\\[10pt]
&\beta_{i} = \frac{\myavtime{i}}{\mystdtime{i}^2}
\end{cases}
\end{equation}
Note that the specific model of user behavior employed in the performance analysis is however not critical, and a similar study could be conducted with other behavior models. 

\begin{table}[!h]
\centering
\center
\subtable[\label{tab:tm}VSM transition matrix.]{
\begin{tabular}{|c|c|c|c|c|}
\hline
Views& \#0& \#1& \#2 &\#3\\
\hline
\#0 &0 &0.4 &0.2 &0.4\\
\hline
\#1 &0.4 &0 &0.4 &0.2\\
\hline
\#2 &0.2 &0.4 &0 &0.4\\
\hline
\#3 &0.4 &0.2 &0.4 &0\\
\hline
\end{tabular}
}
\subtable[\label{tab:sd}VSM state duration parameters.]
{
\begin{tabular}{|c|c|c|}
\hline
Views&av. time& st. dev.\\
\hline
\#0& 6min &30s\\
\hline
\#1 to \#3& 1min &10s\\
\hline
\end{tabular}
}
\caption{\label{tab:swmat} View switching model (VSM) parameters.}
\end{table}

\subsection{Performance analysis}
We compare the traffic load due to the H.264 MVC source and the synthetic video traffic trace generated by our P-HMM model in both network scenarios defined above.

We consider that the MVC traffic is fed into the transmission buffer \mybuft that is characterized by a buffer size \mybuftsize and an output rate \mycr\ (see Figure \ref{fig:sys}). The transmission buffer adopts a First In First Out (FIFO) scheduling policy. The  buffer output is encapsulated into networks packets in accordance with network packetization rules and transmitted to the destination through the channel. Each packet might be affected by a different (random) delay during transmission. The delay $\myovdel{n}$ is the sum of  the channel delay $\mycdel{n}$  and the transmission buffer  delay $\mytdel{n}$. For modeling the channel delay $\mycdel{n}$, we resort to the  quite general and complete channel model introduced by Miao and Chou%
\footnote{ Specifically, we have adopted the same numerical channel model parameters as in \cite{bib:miao},  $\alpha=80,\ n=4,\ \chi=0.025$.} \cite{bib:miao}.
After an initial prefetch delay \mydelay\ (namely \mydelay = 2 sec. in our study) from the arrival time of the first frame, the playout buffer is drained at a rate given by the MVC compressed stream. If frames are not available in the playout buffer at their decoding deadline, they are considered as lost. We consider three different values for the channel rate, namely 1, 1.5, or 2 times  the average bitrate of the  MVC source rate. We denote the ratio between the channel rate and then average source rate by the factor \mychr.
\renewcommand{\hlength}{8cm}
\begin{figure*}[t!]
\center
\subfigure[Transmission buffer\label{fig:3dtvq40tx}]{\includegraphics[width =\hlength, keepaspectratio]{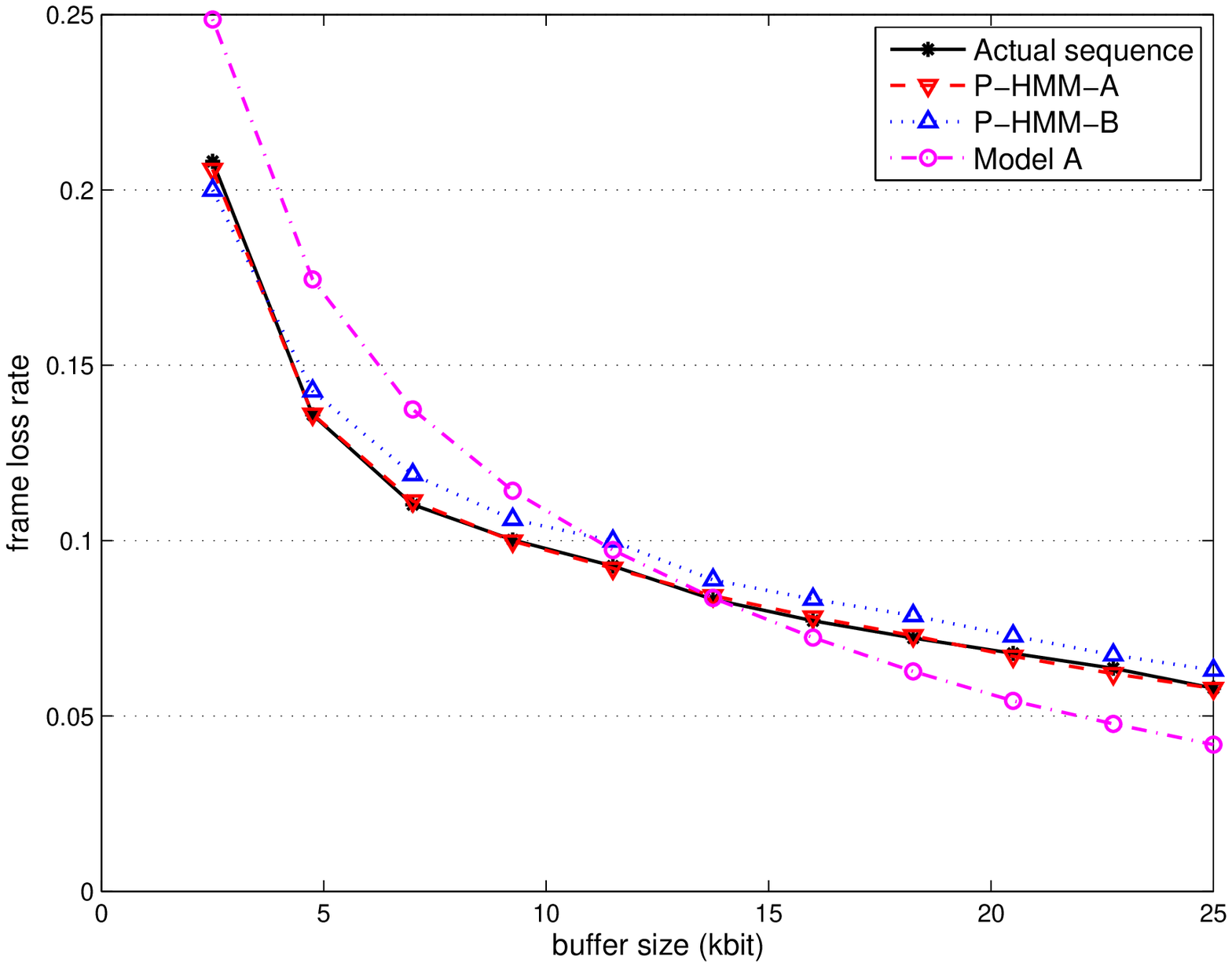}}
\subfigure[Playout buffer\label{fig:3dtvq40rx}]{\includegraphics[width =\hlength, keepaspectratio]{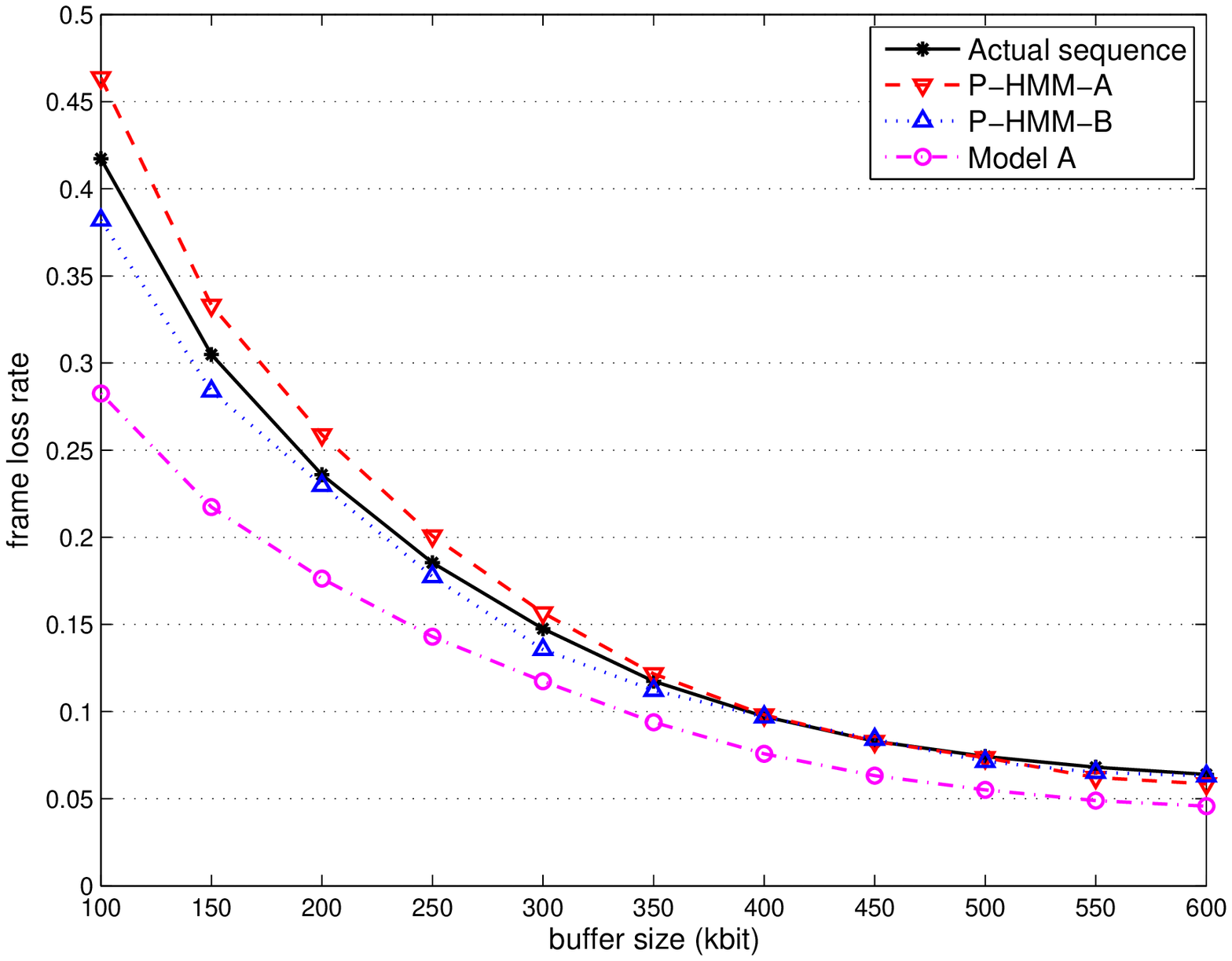}}
\caption{\label{fig:3dtvq40} Frame loss rate in Multiview TV services (low quality stream, \mychr = 2).}
\end{figure*}

We have generated a  25 minute long multiview test sequence, by concatenating the streams described in Table \ref{tab:seq}, similarly to the sequences used in Section \ref{sec:simset}. A 3-state model is built by running the EM algorithm on the actual sequence using the same procedure as Section \ref{sec:simset}. For each sequence, we generated two streams with high (Q=10) and low (Q=40) quality, respectively. We then study the accuracy of our model by comparing the traffic and more particularly the loss rate due to late packets, for both the synthetic traffic and the actual MVC sequence. The loss rate is defined as the ratio between the number of lost frames and the number of  transmitted frames at both the sender and receiver buffers. The frame loss rate is averaged over 10 Monte Carlo simulations.

First, we compare the sender buffer frame loss rate for different values of the buffer size. Figures \ref{fig:3dtvq40tx}, 
 \ref{fig:3dtvq10tx} and Figures
 \ref{fig:inttvq40tx}, \ref{fig:inttvq10tx}
 show respectively these results for the Multiview TV and the Interactive TV cases,
 at different channel rates and stream quality values.
It can be seen that the actual MVC sequence and the synthetic sequence share a similar frame loss rate for different channel rates and transmission buffer sizes. Note that the close similarity is due to the model's capability to describe higher order statistics by means of the non-stationary activity level chain.

Finally, we compare the overall frame loss rate, i.e., the sum of lost frames at both the sender and receiver buffers, divided by the total number of frames, as a function of the receiver buffer size\footnote{To determine the overall frame loss rate, we set the transmission buffer size to be large enough to guarantee that the frame loss rate at the transmission side is not higher than 5\%.}.  Figures \ref{fig:3dtvq40rx}, \ref{fig:3dtvq10rx} and Figures 
 \ref{fig:inttvq40rx}, \ref{fig:inttvq10rx} 
 show these results for the Multiview TV and Interactive TV cases, respectively. Table \ref{tab:flr} summarizes these results for other test settings, quantifying the model's accuracy as the average absolute difference $e$ of the frame loss rate, between the real sequence and the synthetic sequence.

The average is taken over the different buffer sizes under consideration and the frame loss rate is expressed in percent. It can be seen that the synthetic sequence  closely follows the behavior of the actual MVC source; specifically, the difference between the frame loss rate of the model and the one of the actual sequence is smaller than 0.03 for most of the playout buffer sizes under examination.
Since the frame arrival time at the play-out buffer depends on the size of the previous frames that are transmitted, the close similarity between the synthetic and the actual traffic demonstrates the accuracy of the model in characterizing MVC traffic statistics. In addition, even the P-HMM-B model, which is trained on different sequences than the test sequences, is able to capture relevant traffic features of actual data thus providing a good frame loss rate estimation for both transmission and play-out buffer, as seen from Table \ref{tab:flr}.
Therefore, the P-HMM can replace real MVC sequences in the dimensioning of transmit and receive buffers. It can be employed both for synthetic trace generation as well as for theoretical network performance analysis.
\begin{table*}[t]
\center
\subtable[High quality sequence]{
\begin{tabular}{|c|c|c||c|c||c|c|}
\hline
&\multicolumn{2}{|c||}{\mychr = 1}&\multicolumn{2}{|c||}{\mychr = 1.5}&\multicolumn{2}{|c|}{\mychr = 2}\\
\hline
{\bf buffer}&sender&receiver&sender&receiver&sender&receiver\\
\hline
P-HHM-A&0.37&0.5&0.41&0.39&0.15&0.39\\
\hline
P-HHM-B&0.56&1.47&0.7&1.17&0.63&1.06\\
\hline
Model A&1.58&1.84&1.42&1.23&1.11&0.83\\
\hline
\end{tabular}}
\subtable[Low  quality sequence]{
\begin{tabular}{|c|c|c||c|c||c|c|}
\hline
&\multicolumn{2}{|c||}{\mychr = 1}&\multicolumn{2}{|c||}{\mychr = 1.5}&\multicolumn{2}{|c|}{\mychr = 2}\\
\hline
{\bf buffer}&sender&receiver&sender&receiver&sender&receiver\\
\hline
P-HHM-A&0.18&0.39&0.11&0.87&0.09&1.27\\
\hline
P-HHM-B&0.46&1.13&0.61&1.66&0.62&0.87\\
\hline
Model A&1.47&5.68&1.61&5.52&1.68&4.33\\
\hline
\end{tabular}}
\caption{ \label{tab:flr}Average frame loss rate divergence between real sequence and synthetic sequences (Multiview TV case).}
\end{table*}

\begin{table*}[t]
\center
\subtable[High quality sequence]{
\begin{tabular}{|c|c|c||c|c||c|c|}
\hline
&\multicolumn{2}{|c||}{\mychr = 1}&\multicolumn{2}{|c||}{\mychr = 1.5}&\multicolumn{2}{|c|}{\mychr = 2}\\
\hline
{\bf buffer}&sender&receiver&sender&receiver&sender&receiver\\
\hline
P-HHM-A&0.29&0.23&0.32&0.18&0.12&0.23\\
\hline
P-HHM-B&1.08&0.88&1.18&0.75&0.65&0.72\\
\hline
Model A&2.43&1.83&1.69&0.55&1.05&0.86\\
\hline
\end{tabular}}
\subtable[Low  quality sequence]{
\begin{tabular}{|c|c|c||c|c||c|c|}
\hline
&\multicolumn{2}{|c||}{\mychr = 1}&\multicolumn{2}{|c||}{\mychr = 1.5}&\multicolumn{2}{|c|}{\mychr = 2}\\
\hline
{\bf buffer}&sender&receiver&sender&receiver&sender&receiver\\
\hline
P-HHM-A&0.39&1.21&0.34&0.91&0.34&0.91\\
\hline
P-HHM-B&0.59&3.53&0.65&1.50&0.67&1.61\\
\hline
Model A&4.02&10.24&4.22&3.85&4.27&3.94\\
\hline
\end{tabular}}
\caption{ \label{tab:flr}Average frame loss rate divergence between real sequence and synthetic sequences (Interactive TV case).}
\end{table*}

\section{Conclusion}
\label{sec:conc}
In this paper, we have presented a new stochastic model characterizing the frame size sequence for MVC  variable bit rate (VBR) sources.  The model exploits a Poisson Hidden Markov Model representing the random frame sizes of the different MVC encoded views as a function of the random real video scene activity variations. We have also derived a stable EM algorithm that is applicable to long data sequences for the P-HMM parameter estimation. We have shown through extensive simulations that our model accurately predicts the sequence of frame sizes in an MVC stream. We have also applied our model to traffic load prediction in two different network scenarios, namely a multiview TV service and an interactive TV service. Simulation results show that the synthetic traffic generated by the proposed model strongly resembles the traffic due to real MVC video traces. The model is able to accurately characterize a class of MVC streams sharing similar content characteristics with the training data. The model is therefore an appropriate tool for different networking problems, such as network dimensioning, resource allocation, and call admission control.

\renewcommand{\hlength}{8cm}
\begin{figure*}[!htb]
\center
\subfigure[Sender buffer\label{fig:3dtvq10tx}]{\includegraphics[width =\hlength, keepaspectratio]{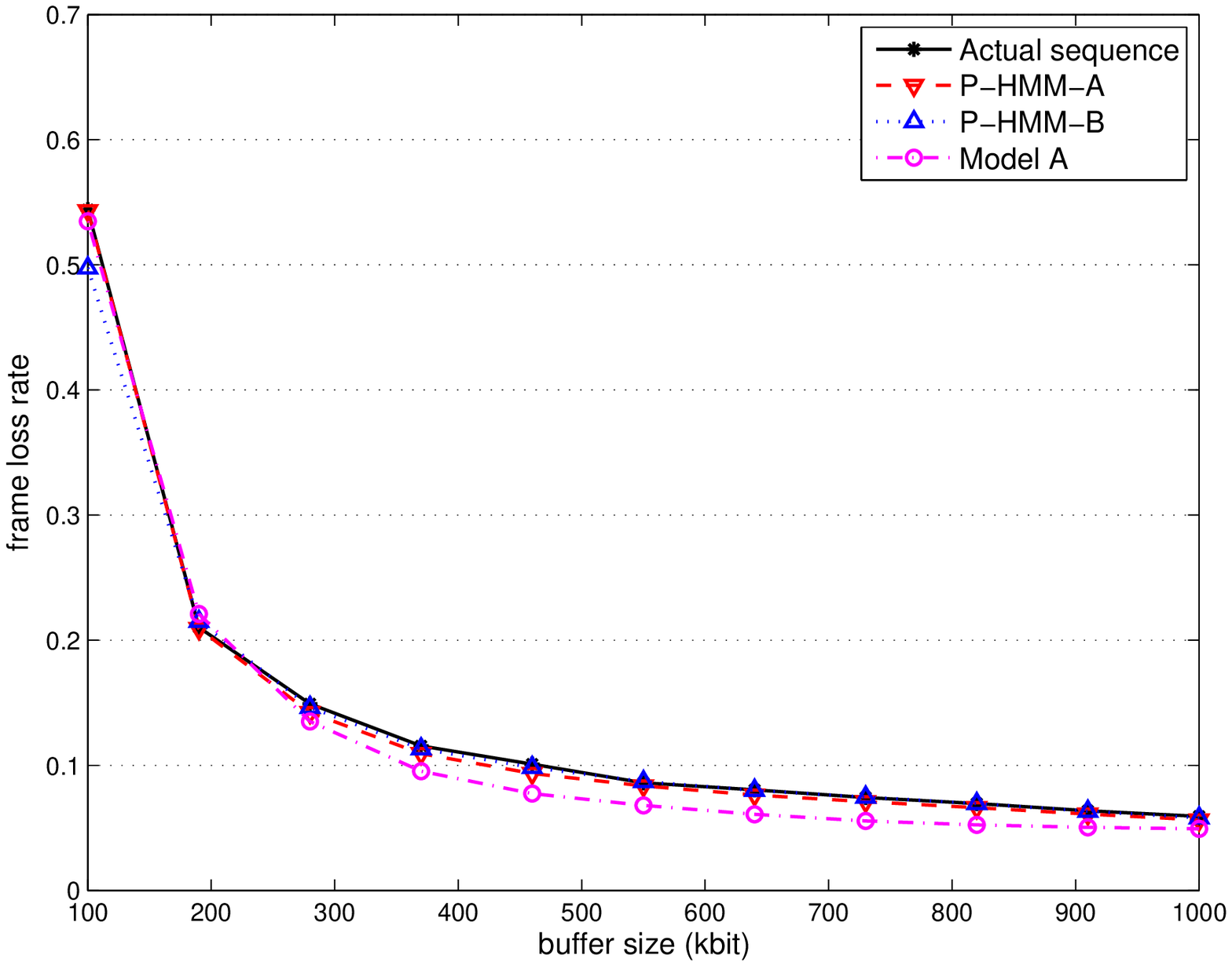}}
\subfigure[Receiver buffer\label{fig:3dtvq10rx}]{\includegraphics[width =\hlength, keepaspectratio]{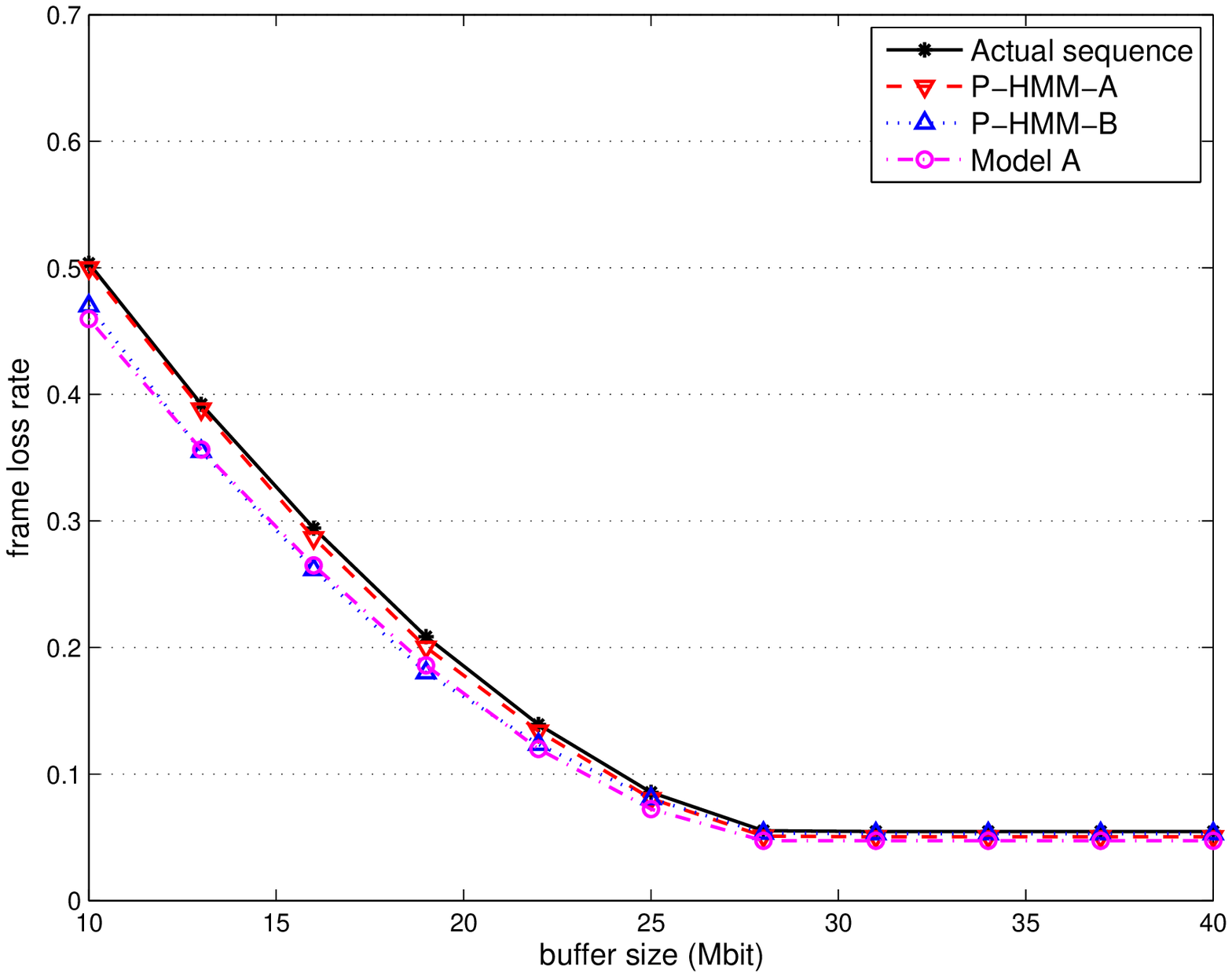}}
\caption{\label{fig:3dtvq10} Frame loss rate,  Multiview TV services (high quality stream, \mychr = 1).}
\end{figure*}

\renewcommand{\hlength}{8cm}
\begin{figure*}[!tb]
\center
\subfigure[Sender buffer\label{fig:inttvq40tx}]{\includegraphics[width =\hlength, keepaspectratio]{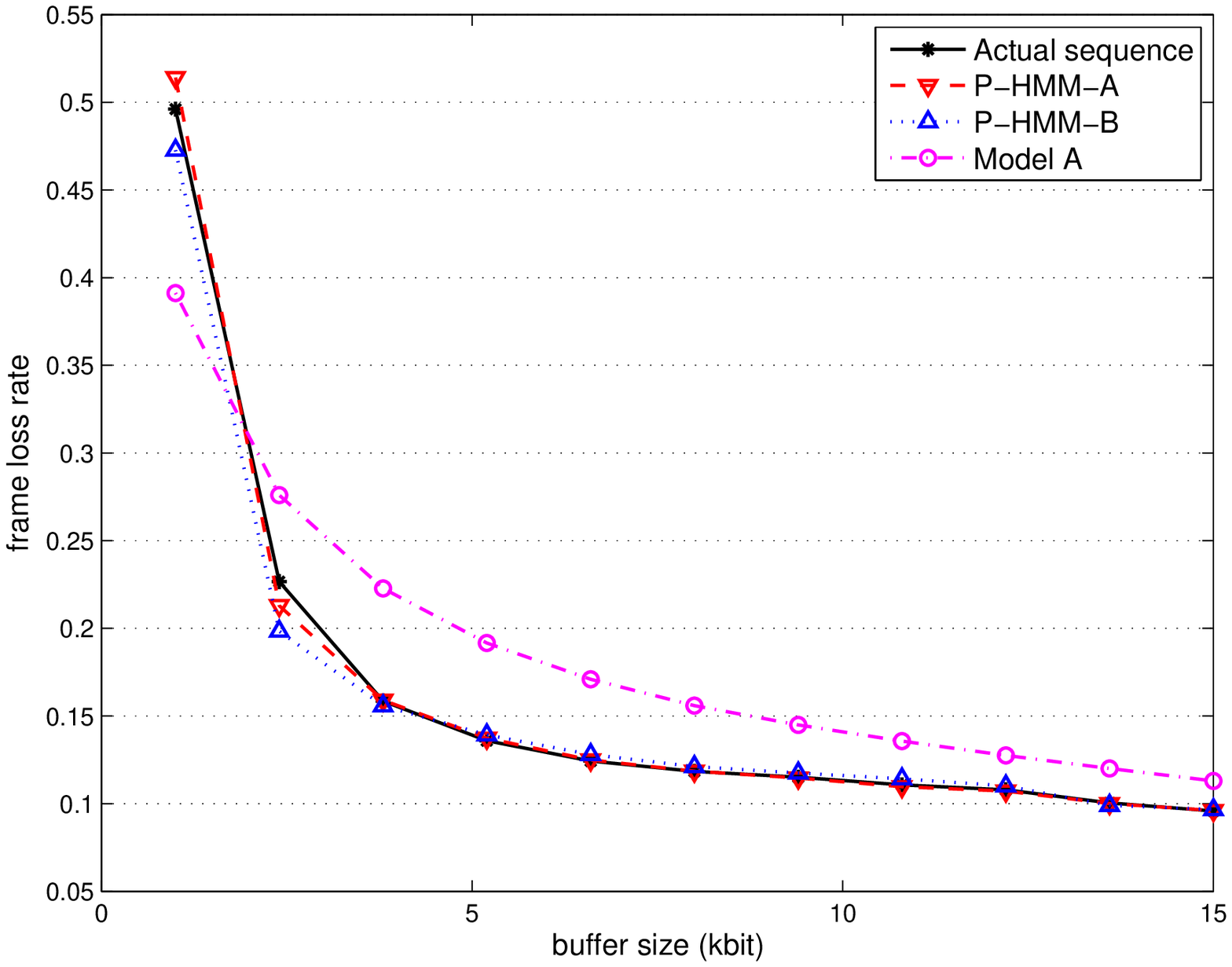}}
\subfigure[Receiver buffer\label{fig:inttvq40rx}]{\includegraphics[width =\hlength, keepaspectratio]{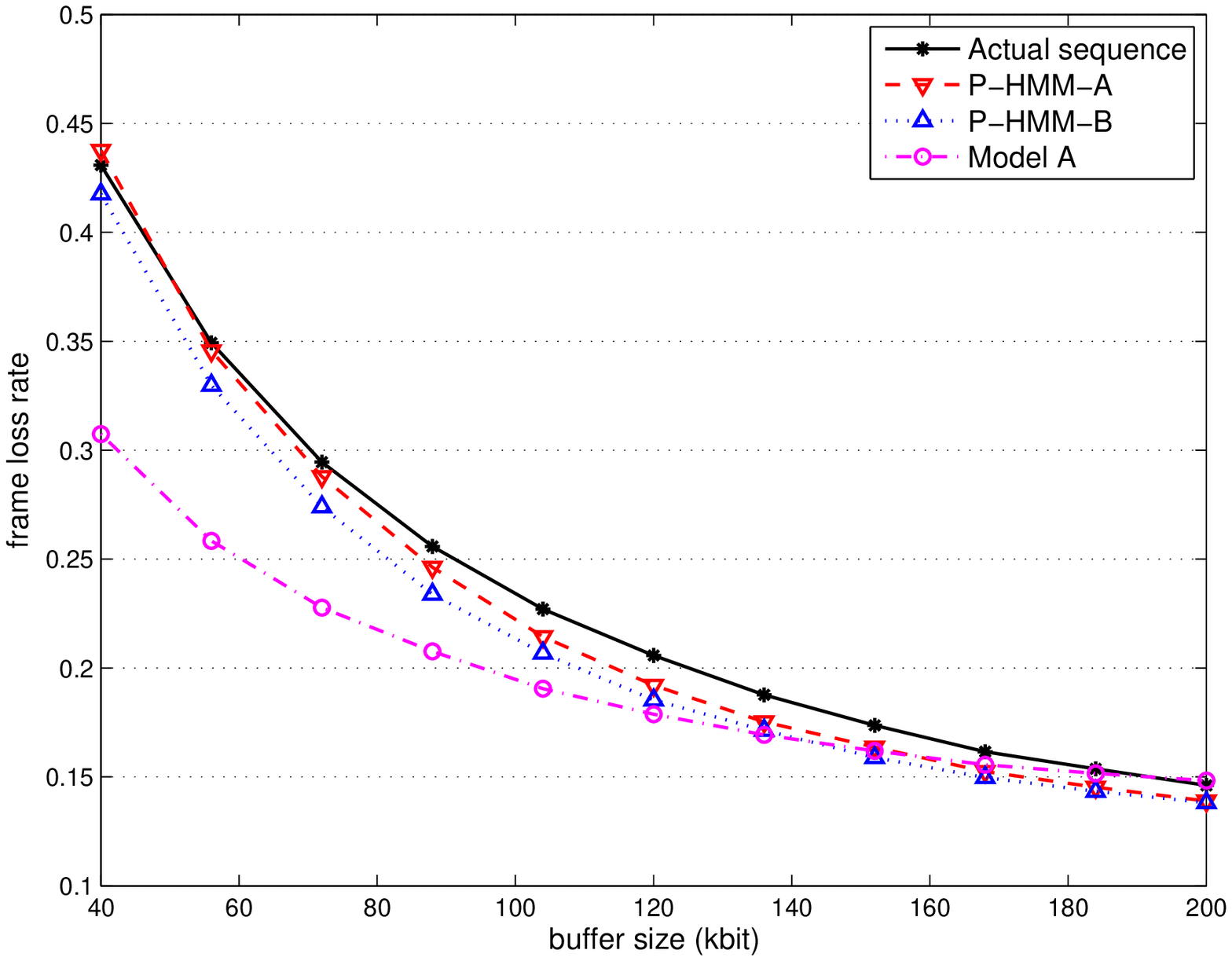}}
\caption{\label{fig:inttvq40} Frame loss rate in Interactive TV services(low quality stream, \mychr = 2).}
\end{figure*}

\renewcommand{\hlength}{8cm}
\begin{figure*}[!htb]
\center
\subfigure[Sender buffer\label{fig:inttvq10tx}]{\includegraphics[width =\hlength, keepaspectratio]{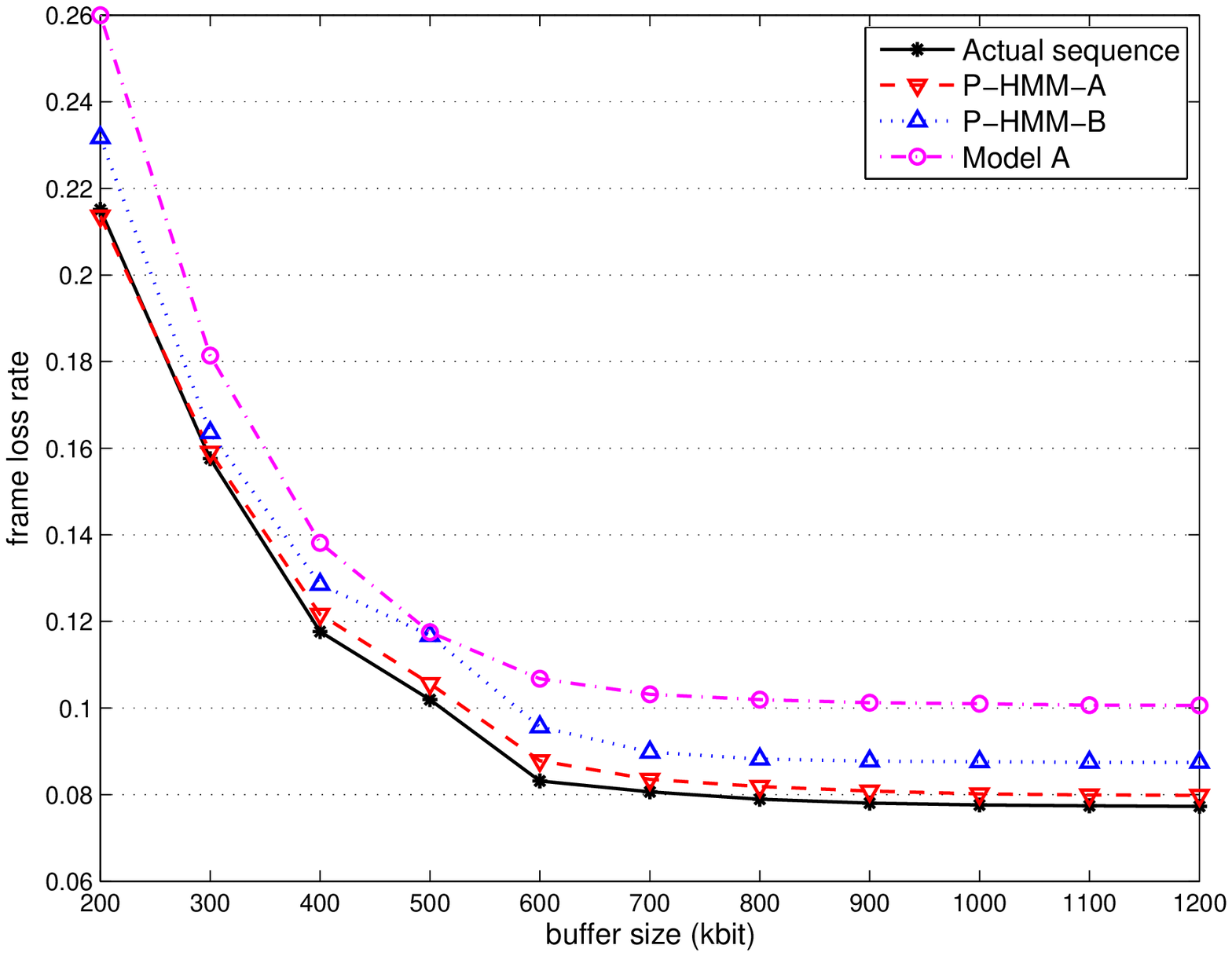}}
\subfigure[Receiver buffer\label{fig:inttvq10rx}]{\includegraphics[width =\hlength, keepaspectratio]{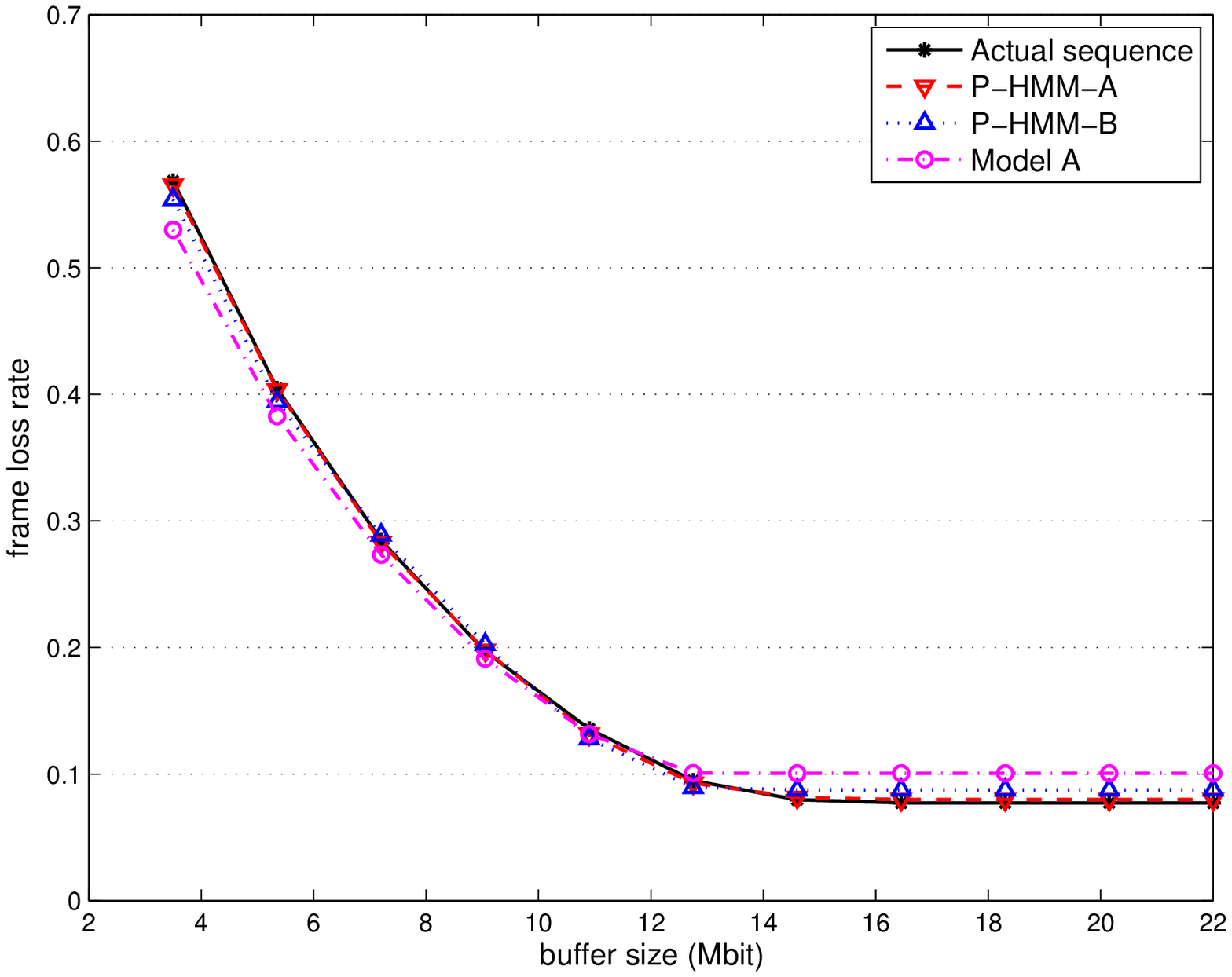}}
\caption{\label{fig:inttvq10} Frame loss rate in Interactive TV services (high quality stream, \mychr = 1).}
\end{figure*}

\end{document}